\documentclass[%
 reprint,
groupedaddress,
amsmath,amssymb,
aps,
prl,
]{revtex4-1}
\usepackage{graphicx}% Include figure files
\usepackage{dcolumn}% Align table columns on decimal point
\usepackage{bm}% bold math
\usepackage{float}
\usepackage{multirow}
\usepackage[dvipdfm,colorlinks,linkcolor=blue, urlcolor=blue, anchorcolor=blue, citecolor=blue]{hyperref}
\begin{document}

%\preprint{APS/123-QED}

\title{Wireless energy transfer in non-Hermitian quantum battery}% Force line breaks with \\
\author{Fang-Mei Yang$^{1}$}%
\author{Fu-Quan Dou$^{1,2,}$}
\email{doufq@nwnu.edu.cn}
\affiliation{$^{1}$College of Physics and Electronic Engineering, Northwest Normal University, Lanzhou, 730070, China
\\
$^{2}$Gansu Provincial Research Center for Basic Disciplines of Quantum Physics, Lanzhou, 730000, China}
\begin{abstract}
The extraction of energy is one of fundamental challenges in realizing quantum batteries (QBs). Here, we propose two wireless transfer schemes with parity-time symmetries to efficiently extract the energy stored in non-Hermitian QBs to consumption centers. For linear cases, the transfer energy oscillates periodically in the unbroken symmetry region and grows hyperbolically in the broken region. For nonlinear cases, the transfer energy eventually reach and remain steady-state values arising from the feedback mechanism of the nonlinear saturable gain. Furthermore, we show the significant robustness and the ultrafast response of the wireless transfer schemes to sudden movements around one metre. Our work overcomes energy bottlenecks for wireless transfer schemes in QBs and may provide inspirations for practical applications of QBs.
%shed light on the explorations of desirable functionality in fundamental research and practical applications.
\end{abstract}
\maketitle
%\tableofcontents
\emph{Introduction.}$-$With the growing attention of energy innovations and the ongoing miniaturization of electronic devices, the applications of quantum engineering on energy technologies are fast emerging and are expected to make substantial progresses in the subdomains of quantum solar, quantum nuclear and quantum battery (QB) \cite{Menke2014,Haber2017,quach2020,metzler2023quantum,Li2018}.
Quantum batteries (QBs) are quantum systems with discrete energetic states, which are deliberately excited to temporarily store energy \cite{PhysRevE.87.042123,RevModPhys.96.031001}. Compared with classical batteries, QBs can break speed limitations of the charging-discharging dynamics and exhibit significant speedup associated with superabsorption and superradiance mechanisms via entangling unitary operations \cite{quach2020,metzler2023quantum,Li2018,PhysRevE.87.042123,RevModPhys.96.031001,PhysRevB.99.205437}. In the thermodynamic limit, it was shown that the superlinear scaling relation between the maximum power and the number of cells in QBs, which far exceeds the classically achievable linear scaling \cite{PhysRevLett.120.117702,PhysRevLett.128.140501,PhysRevLett.125.236402,PhysRevLett.127.100601,PhysRevLett.118.150601,PhysRevB.105.115405,PhysRevA.103.052220,PhysRevA.105.062203,PhysRevA.106.032212,PhysRevE.103.042118}.

Since when 2013 Alicki and Fannes proposed the concept of QB, significant efforts have been paid on innovating models, optimizing performances and exploring implementation schemes \cite{PhysRevA.97.022106,PhysRevA.104.032606,PhysRevE.100.032107,PhysRevA.110.032205,PhysRevB.109.235432,PhysRevLett.122.210601,PhysRevB.99.035421,PhysRevA.103.033715,Shaghaghi2022,PhysRevLett.132.210402,PhysRevResearch.4.013172,PhysRevA.100.043833,PhysRevLett.124.130601,PhysRevA.102.052223,PhysRevLett.132.090401,PhysRevA.109.042207,Dou2020,PhysRevA.109.032201,PhysRevA.107.023725,Cruz2022,quach2020,gemme2022ibm,PhysRevA.106.042601,PhysRevLett.131.260401,PhysRevLett.131.240401,Hu2022,Zheng2022,hymas2025experimental}. These early theoretical and experimental explorations have uncovered some prominent advantages of QBs, ranging from ultrasmall size, ultrafast charging, ultralarge capacity, to ultraslow aging \cite{PhysRevB.102.245407,Rosa2020,PhysRevApplied.14.024092,PhysRevA.102.060201,PhysRevLett.131.030402,PhysRevE.104.024129,PhysRevA.104.032207,PhysRevA.104.042209,PhysRevA.109.022607,PhysRevA.111.012212,PhysRevA.109.062432,PhysRevLett.122.047702,PhysRevLett.129.130602,PhysRevA.109.052206,Dou2021}. Meanwhile, several researchers predicted that QBs may find key uses in future fusion power plants, electric vehicles and medical devices \cite{metzler2023quantum,PhysRevLett.128.140501,RevModPhys.96.031001}. However, the development of QBs, which is still in its infancy now \cite{RevModPhys.96.031001,RevModPhys.79.135,RevModPhys.90.015002,RevModPhys.86.153,RevModPhys.93.025005,RevModPhys.89.035002,RevModPhys.74.145}, mean that the application of QBs remains extremely challenging. One pressing challenge is the way to transfer the energy stored in QBs to consumption centers.

Symmetries are the most fundamental properties of nature, which are responsible for various physical phenomena \cite{RevModPhys.65.569,RevModPhys.88.035005,RevModPhys.88.035002}. Back in 1998, Bender and Boettcher indicated that parity $(\mathcal{P})$ and time $(\mathcal{T})$ symmetries can account for real energy spectra appearing in non-Hermitian systems \cite{PhysRevLett.80.5243}. Subsequently, non-Hermitian systems with $\mathcal{PT}$ symmetries attracted considerable attention \cite{PhysRevLett.80.5243,PhysRevLett.100.030402,PhysRevA.83.041805,PhysRevA.82.013629,Feng2017,PhysRevLett.86.787,Gao2015,Xiao2017,Li2019,El-Ganainy2018,Ashida2020,RevModPhys.88.035002,RevModPhys.93.015005,Ruter2010,PhysRevLett.117.123601} and stimulated many novel applications, such as lasing, sensing, topological insulator, unidirectional invisibility and nonreciprocal light propagation \cite{Peng2014,Harari2018,Hokmabadi2019,PhysRevLett.121.073901,Zhang2024,Mao2024,Chen2017}.
In addition, it was also demonstrated that adiabatic topological operation allows for nonreciprocal energy transfer between two vibrational modes of an optomechanical system with exceptional points \cite{Xu2016}, while a $\mathcal{PT}$-symmetric LRC circuit incorporating nonlinear saturable gain elements provides robust wireless power transfer with variable transfer distance between source and receiver resonators \cite{Assawaworrarit2017}. Inspired by such counterintuitive advantages of energy transfer in classical non-Hermitian systems, it is natural to wonder the possibility of using $\mathcal{PT}$-symmetric theories to control energy distribution in QBs.

In this work, we concentrate on how to transfer the energy stored in QBs to consumption centers and propose two wireless transfer schemes in non-Hermitian QBs. We explore the dynamics of non-Hermitian QBs with linear or nonlinear $\mathcal{PT}$ symmetries and the effects of the variable separation distance between the battery and the consumption-hub on the wireless transfer schemes. To physically describe the linear or nonlinear gain and the magnetic coupling in wireless transfer schemes, we introduce a classical simulator consisting of two magnetically coupled LRC circuits and simulate some counterintuitive advantages of non-Hermitian QBs with $\mathcal{PT}$ symmetries in Supplemental Material \cite{supplementalmaterial}.

\begin{figure}[htbp]
\centering
\includegraphics[width=0.47\textwidth]{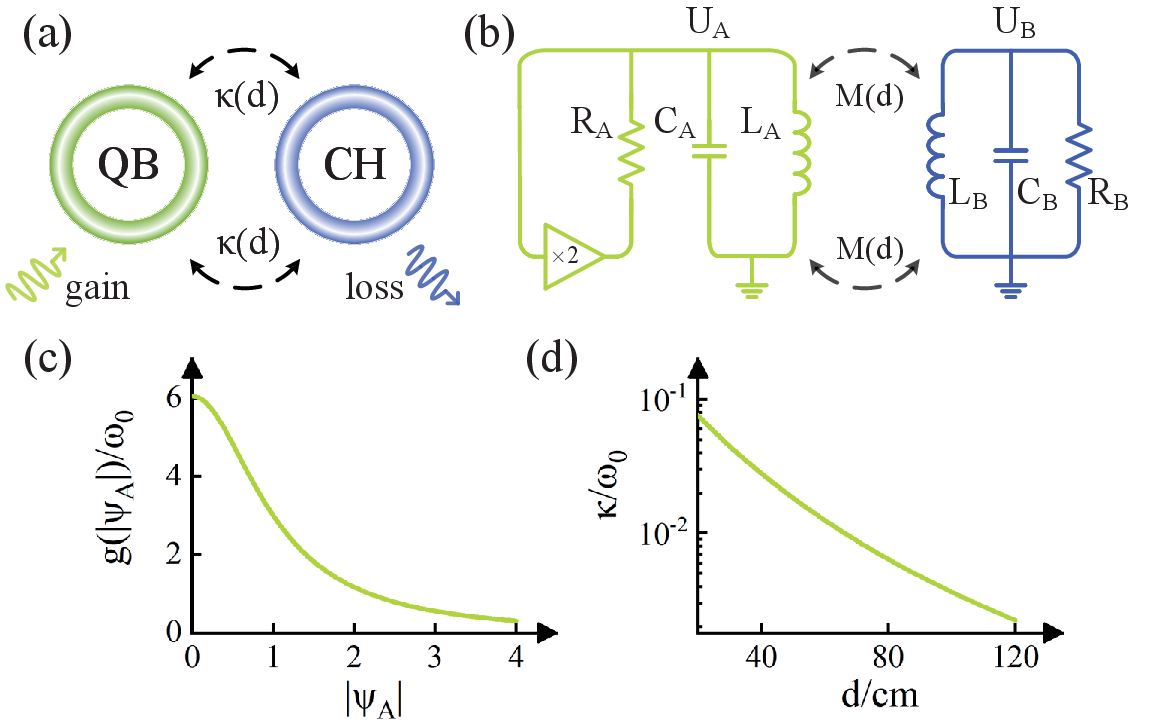}
\caption{(a) The schematic diagram of the non-Hermitian QB with wireless transfer scheme, in which two resonators acts as the battery and the consumption-hub (CH). (b) The classical simulator consisting of two magnetically coupled LRC circuits is used to verify $\mathcal{PT}$ symmetries in non-Hermitian QBs. The effective negative resistor is provided by a voltage-doubling buffer $(\times2)$. (c) The nonlinear gain as a function of the field amplitude in left resonator, where the intrinsic loss rate and gain rate are $\gamma_{1}=0.05\omega_{0},~g_{1}=3\omega_{0}$. (d) The dependence of the coupling strength on the separation distance between two coaxially aligned coils \cite{Assawaworrarit2017}. The resonant frequency is taken as $\omega_{0}=1$.}
\label{fig1}
\end{figure}

\emph{Model.}$-$We consider a non-Hermitian QB with wireless transfer scheme as schematically depicted in Fig.~\ref{fig1}~(a), which consists of two separate quantum resonators acting as the battery and the consumption-hub. The system dynamics is described by the equations ($\hbar=1$)
\begin{equation}
\label{Sequation}
  i\frac{d}{dt}
  \begin{pmatrix}
  \psi_{A}\\\psi_{B}
  \end{pmatrix}
  =H
  \begin{pmatrix}
  \psi_{A}\\\psi_{B}
  \end{pmatrix},
\end{equation}
and the Hamiltonian follows as
\begin{equation}
\label{Ham}
H=\begin{pmatrix}
  \omega_{A}+ig(|\psi_{A}|)&\kappa(d)\\
  \kappa(d)&\omega_{B}-i\gamma
  \end{pmatrix},
\end{equation}
where $\omega_{A}~(\omega_{B})$ is the frequency of the left (right) resonator, $\psi_{A,B}$ represents the corresponding field amplitude for each resonator, the coupling strength $\kappa$ varies as a function of the separation distance $d$ between two resonators, the gain $g$ depends on the field amplitude of the left resonator, and $\gamma$ denotes the loss contributed by the right resonator's load.

Nowadays, many counterintuitive phenomena of non-Hermitian systems have been observed on quantum and classical simulators, including optical waveguides, microwave cavities, atomic lattices, nitrogen-vacancy centers, superconducting circuits and electronic circuits \cite{Ruter2010,Peng2014,PhysRevLett.117.123601,PhysRevLett.126.170506,Naghiloo2019,PhysRevA.84.040101,Assawaworrarit2017}.
In contrast to quantum simulation, classical simulation is more popular because of its low cost, mature technology and strong scalability \cite{Buluta2009,PRXQuantum.4.020340,Ashida2020}. As shown in Fig.~\ref{fig1}~(b), we introduce a simple electronic circuit consisting of two magnetically coupled LC resonators with gain and loss, which are realized by the effective negative resistor $-R_{A}$ and the normal resistor $R_{B}$, respectively. The nonlinear gain function satisfies the form
\begin{equation}
\label{gain}
g(|\psi_{A}|)=\frac{2(g_{1}+\gamma_{1})}{1+|\psi_{A}|^2}-\gamma_{1},
\end{equation}
where $\gamma_{1}$ and $g_{1}$ are the intrinsic loss rate and gain rate of the left resonator. In particular, the linear gain corresponds to the case where the above gain function is constant. Without direct interaction, the coupling between two separate resonators is regulated by the magnetic induction of coils wound into inductors. In the quasi-static limit, the coupling strength $\kappa$ between two identical copper coils spaced coaxially is related to the mutual inductance and self inductance
\begin{equation}
\kappa(d)=\frac{\omega_{0}}{2}\frac{M(d)}{L},
\end{equation}
where $\omega_{0}=\omega_{A}=\omega_{B}$ represents the resonant frequency and the mutual (self) inductance $M(L)$ can be calculated in Ref.~\cite{Assawaworrarit2017}. The derivations of electronic circuits and the observations of $\mathcal{PT}$ symmetries in non-Hermitian QBs are presented in Supplemental Material \cite{supplementalmaterial}.

The initial states of the battery and the consumption-hub are prepared to the single-photon state $|1\rangle$ and the vacuum state $|0\rangle$, respectively. The energy transfer from the battery to the consumption-hub is defined as the energy stored in the right resonator
\begin{equation}
\label{energy}
E(t)=\omega_{B}|\psi_{B}|^2,
\end{equation}
and the average transfer power is given by
\begin{equation}
P(t)=E(t)/t.
\end{equation}

\emph{Wireless transfer process.}$-$Magnetic field coupling has been extensively developed in the wireless power transfer applications. In the following, we mainly focus on the wireless transfer processes of the non-Hermitian QB using magnetic resonance coupling in the near-field conditions. The characteristic equation of the non-Hermitian system at a specific separation distance $(\kappa=0.5\omega_{0})$ is as follows
\begin{equation}
\label{characteristic_eq}
\left(\omega_{0}+ig-\omega\right)\left(\omega_{0}-i\gamma-\omega\right)-\kappa^{2}=0.
\end{equation}

For a linear gain $g=\gamma$, the Hamiltonian~(\ref{Ham}) commutes with the operator $\mathcal{PT}$ and satisfies $\mathcal{PT}H(\mathcal{PT})^{-1}=H$, where the parity operator $\mathcal{P}$ exchanges the positions of two resonators and the time reversal operator $\mathcal{T}$ performs a complex conjugate operation. The eigenfrequencies of the linear non-Hermitian system are specified by $\omega=\omega_{0}\pm\sqrt{\kappa^{2}-\gamma^{2}}$, and the energy defined in Eq.~(\ref{energy}) can be obtained by solving Eq.~(\ref{Sequation})
\begin{equation}
\label{analyticenergy}
E=
\begin{cases}
\omega_{0}\kappa^{2}t^{2},&|\gamma|=|\kappa|,\\
\omega_{0}\kappa^{2}\sin^{2}(\sqrt{\kappa^{2}-\gamma^{2}}t)/(\kappa^{2}-\gamma^{2}),&|\gamma|<|\kappa|,\\
\omega_{0}\kappa^{2}\sinh^{2}(\sqrt{\gamma^{2}-\kappa^{2}}t)/(\gamma^{2}-\kappa^{2}),&|\gamma|>|\kappa|.
\end{cases}
\end{equation}

\begin{figure}[htbp]
\centering
\includegraphics[width=0.47\textwidth]{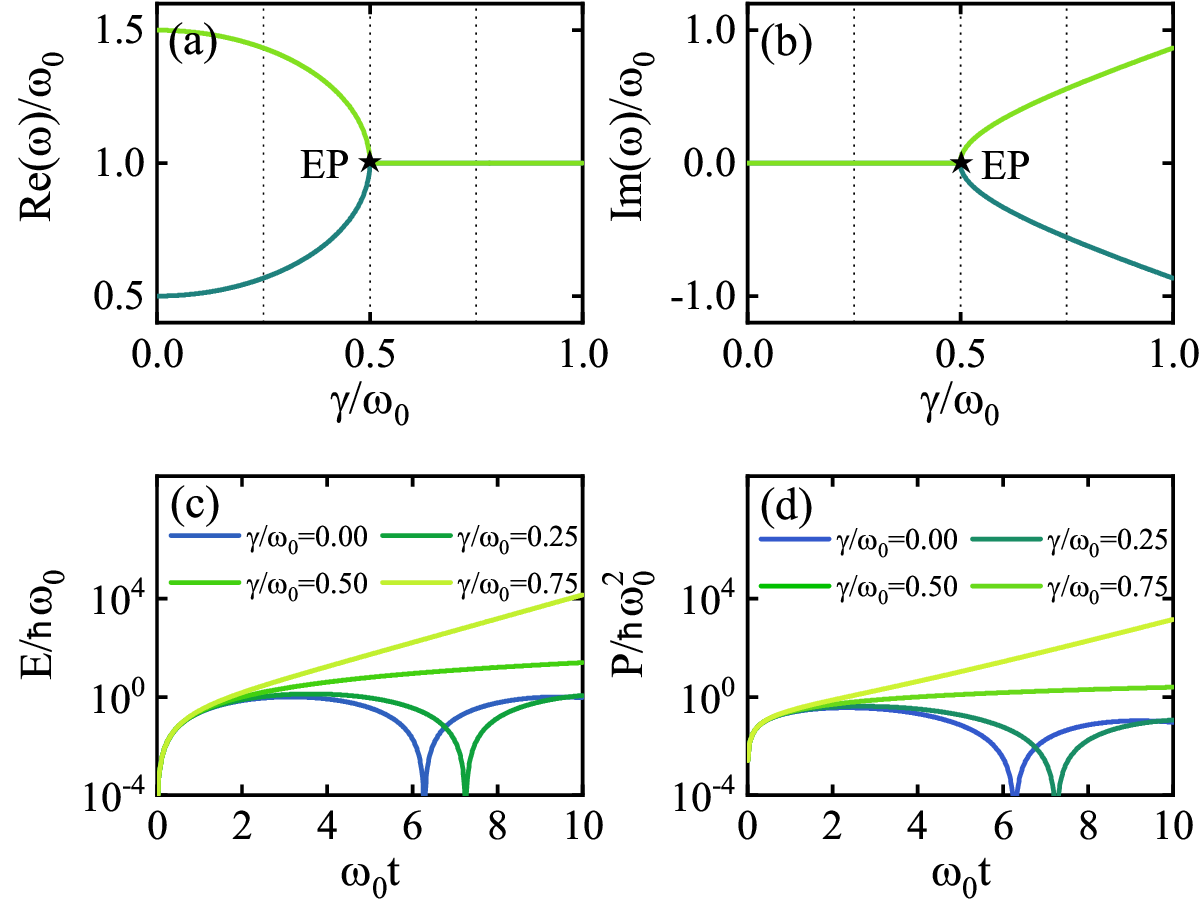}
\caption{The non-Hermitian QB with linear gain.~(a),~(b) The real and imaginary parts of the eigenfrequencies $\omega$ in dependence on the loss rate $\gamma$. The parameter space around the EP can be divided into unbroken and broken $\mathcal{PT}$-symmetry regions depending on whether the eigenfrequency is purely real. (c),~(d) The time evolution of the energy $E$ and the power $P$ for different values of the loss rate $\gamma$. Because of the nonorthogonality of the eigenvectors, the transfer dynamics shows a pseudo-closed oscillation in the unbroken region and grows exponentially in the broken region.}
\label{fig2}
\end{figure}
It can be observed from the analytic solutions that there is a link between the eigenfrequencies and the performance of the linear non-Hermitian QB. In order to understand the underlying link, we illustrate the real and imaginary parts of the eigenfrequencies as a function of the loss rate and the dynamics of the consumption-hub for different values of the loss rate as shown in Fig.~\ref{fig2}. Intriguingly, the exceptional point (EP), the essence of the unusual properties of non-Hermitian systems \cite{Ashida2020}, appears when eigenfrequencies degenerate and eigenvectors coalesce. The appearance of the EP divides the parameter space of Figs.~\ref{fig2}~(a)~and~(b)~into two regions: unbroken and broken $\mathcal{PT}$ symmetry, which has a crucial impact on the linear non-Hermitian QB. In the unbroken region $|\gamma|<|\kappa|$, two unequal eigenfrequencies remain purely real and the energy shows a Rabi-type oscillation between the battery and the consumption-hub. At the EP for $|\gamma|=|\kappa|$, the two purely real eigenfrequencies become equal. The oscillatory behavior is completely disrupted and the energy increases proportionally with the quadratic power of time. In the broken region $|\gamma|>|\kappa|$, the non-Hermitian system has a pair of complex conjugate eigenfrequencies and the energy grows exponentially with one eigenfrequency while decays exponentially with the other, that is hyperbolic-sinusoidal growth. Overall, the pseudo-closed oscillations or the monotonic growths of energy in the consumption-hub mainly depend on whether the eigenfrequency of the linear non-Hermitian system is purely real. Meanwhile, it is worth mentioning that the energy and the power transferred from the linear non-Hermitian QB are increased by several orders of magnitude than the Hermitian QB described in Figs.~\ref{fig2}~(c) and (d) for $\gamma=0$. The dynamic advantage mainly originates from the nonorthogonality of the eigenvectors caused by the presence of the non-Hermiticity. The results demonstrate that the non-Hermiticity, introduced by linear $\mathcal{PT}$ symmetry, promotes energy transfer while inhibits dynamic oscillation in the wireless transfer processes of QBs.

\begin{figure}[htbp]
\centering
\includegraphics[width=0.47\textwidth]{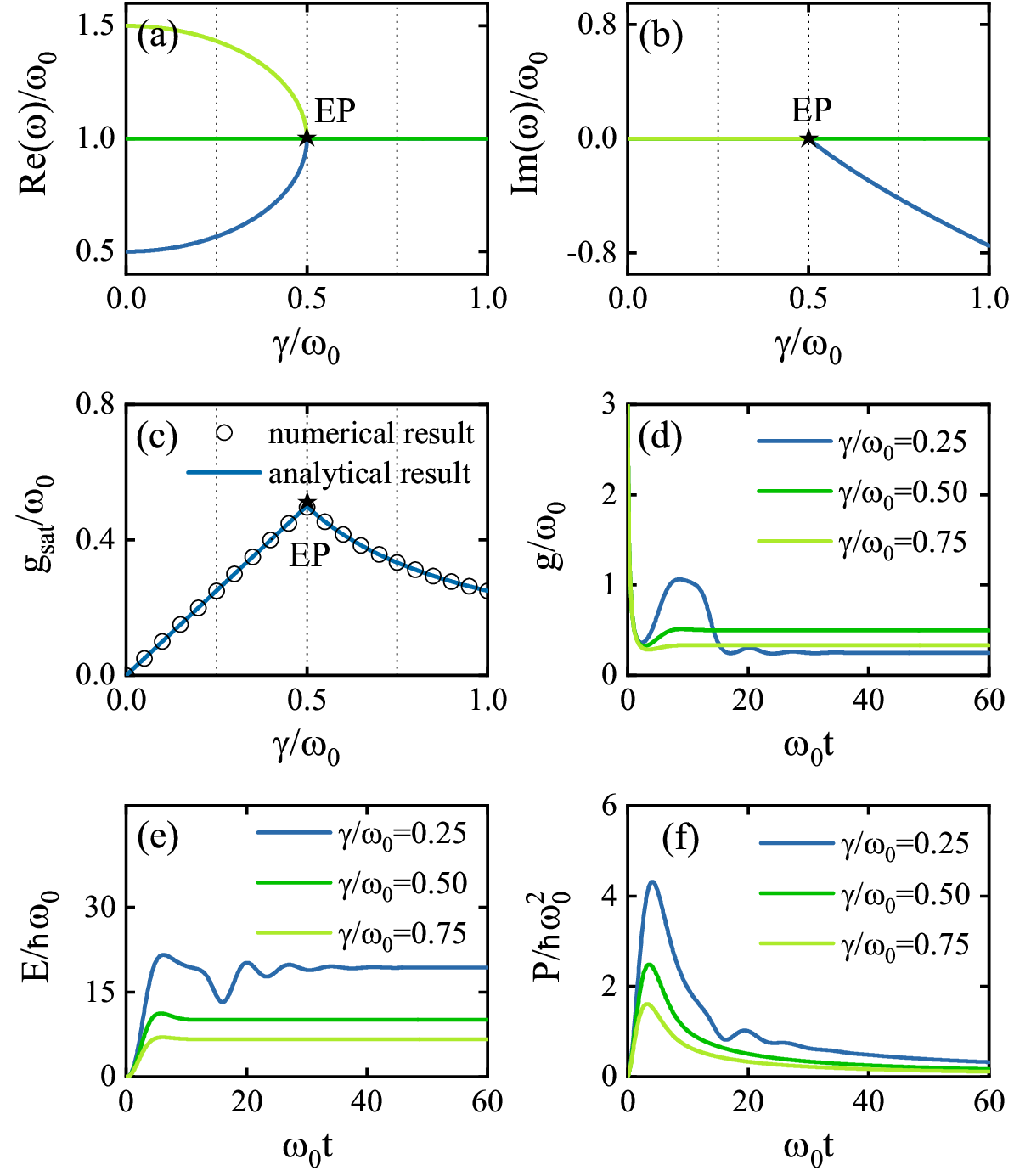}
\caption{The non-Hermitian QB with nonlinear gain.~(a)-(c) The eigenfrequencies $\omega$ and the saturated gain $g_{sat}$ in dependence on the loss rate $\gamma$. The saturated gain increases proportionally in the unbroken region while decreases inversely in the broken region, and the numerical result matches perfectly with the analytical result. (d)-(f) The time evolution of the nonlinear gain $g$ and the dynamics of the consumption-hub for different values of the loss rate $\gamma$. The feedback mechanism of the nonlinear saturable gain causes the consumption-hub to reach and remain at steady states.}
\label{fig3}
\end{figure}
Unlike the approach used for the study of linear non-Hermitian system, then we consider a non-Hermitian system with nonlinear gain shown in Eq.~(\ref{gain}). In the unbroken region, the differences between the resonant frequency $\omega_{0}$ and the eigenfrequencies $\omega$ are purely real. The real and imaginary parts of characteristic equation~(\ref{characteristic_eq}) can be separated to
\begin{equation}
\label{realeq}
\left(\omega_{0}-\omega\right)^{3}-\left(\kappa^{2}-\gamma^{2}\right)\left(\omega_{0}-\omega\right)=0,
\end{equation}
and
\begin{equation}
\label{unbrokensaturatedgain}
g_{sat}=\gamma,
\end{equation}
where the eigenfrequencies $\omega=\omega_{0},~\omega_{0}\pm\sqrt{\kappa^{2}-\gamma^{2}}$. The corresponding saturated gain $g_{sat}$ provides a gain value at which the system oscillates as a steady state with eigenfrequency $\omega$. In the broken region, the frequency differences become complex and the characteristic equation~(\ref{characteristic_eq}) can be written as
\begin{equation}
\begin{aligned}
&{\rm Im}\left[\left(\omega_{0}-\omega\right)^{2}+i\left(g-\gamma\right)\left(\omega_{0}-\omega\right)\right]=0,\\
&{\rm Re}\left[\left(\omega_{0}-\omega\right)^{2}+i\left(g-\gamma\right)\left(\omega_{0}-\omega\right)\right]+g\gamma-\kappa^{2}=0.
\end{aligned}
\end{equation}
According to the continuity of the saturated gain and the eigenfrequencies $\omega=\omega_{0}$ at the EP, we obtain
\begin{equation}
\label{brokensaturatedgain}
g_{sat}=\frac{\kappa^{2}}{\gamma},
\end{equation}
and the corresponding eigenfrequencies in the broken region become as $\omega=\omega_{0},~\omega_{0}+i(\kappa^{2}/\gamma-\gamma)$.

The eigenfrequencies of the nonlinear non-Hermitian system in dependence on the loss rate and the dynamical evolutions of the consumption-hub for different loss rates are shown in Fig.~\ref{fig3}. Similar to the linear non-Hermitian system, the parameter space around the EP can be separated into two regions according to the relative values of the coupling strength and the loss rate. In contrast to the linear system, the nonlinear non-Hermitian system exists three eigenfrequencies in the unbroken region and the consumption-hub eventually reach and remain steady states with considerable energy in the total regions. This originates from the feedback mechanism of the nonlinear saturable gain, which is defined as the nonlinear gain strength corresponding to the steady states of the consumption-hub. In particular, the saturated gain increases proportionally with loss rate in the unbroken region while decreases inversely with loss rate in the broken region, see Figs.~\ref{fig3}~(c),~(d) and Eqs.~(\ref{unbrokensaturatedgain}),~(\ref{brokensaturatedgain}).

\begin{figure}[htbp]
\centering
\includegraphics[width=0.47\textwidth]{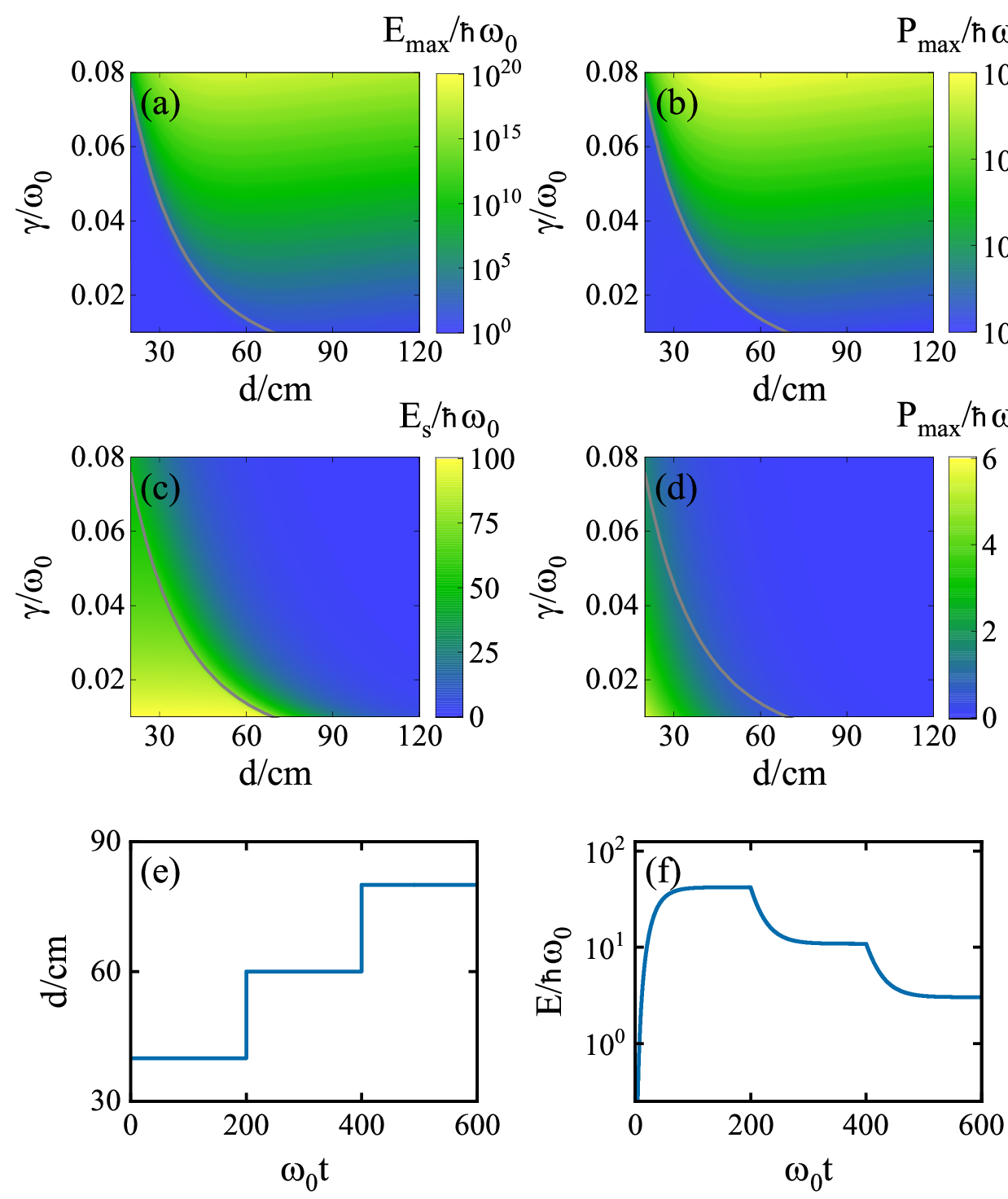}
\caption{The robustness and the transient response of the wireless transfer processes versus the separation distance. (a)-(d) Contour plots of the maximum energy $E_{max}$, the stable energy $E_{s}$ and the maximum power $P_{max}$ as functions of the separation distance $d$ and the loss rate $\gamma$. The left and right sides of the gray curves indicate the unbroken and broken regions, respectively. (e),~(f) The effect of sudden changes in the separation distance $d$ on the energy $E$ transferred from the nonlinear non-Hermitian QB to the consumption-hub, where the loss rate $\gamma=0.04\omega_{0}$. The consumption-hub can settle into steady states within an instant when the separation distance suddenly shifts between $20~{\rm cm}$ to $120~{\rm cm}$.}
\label{fig4}
\end{figure}
The robustness and the transient response versus the variable separation distance between the battery and the consumption-hub are critical in the wireless transfer processes. To reveal how the distance specifically affects the behaviors of the energy and the power transferred from the non-Hermitian QBs, we introduce three physical parameters which characterized by
\begin{equation*}
E_{max}={\rm max}(E),~E_{s}=E(\infty),~P_{max}={\rm max}(P).
\end{equation*}
Figure~\ref{fig4} describes the maximum energy (the stable energy) and the maximum power, transferred from the non-Hermitian QB with linear (nonlinear) gain, as functions of the separation distance and the loss rate. Noteworthily, the parameter space can be distinguished as the unbroken and broken regions by the exceptional arcs~(the gray curves $\gamma=\kappa$)~\cite{Zhou2018,Tang2020}. For the increased loss rate, the maximum energy and the maximum power persistently increase under the linear gain while the stable energy and the maximum power remarkably decrease under the nonlinear gain. This leads to the optimal working regions being completely different in two non-Hermitian QBs, where the linear and nonlinear QBs are recommended to work in the broken and unbroken regions, respectively. For the fixed loss rate, the three physical parameters are almost unchanged with the separation distance no matter the non-Hermitian QB is in the unbroken or broken region. The results mean that the wireless transfer processes of non-Hermitian QBs are significantly robust to the separation distance. In practice, it is extremely difficult to observe any transient response with continuous movements. Thus, we visualize the dynamic wireless transfer processes by manipulating the separation distance variations in the form of step functions. As shown in Figs.~\ref{fig4}~(e) and (f), the energy from the nonlinear non-Hermitian QB can quickly return to steady states with the sudden movements between $20~{\rm cm}$ to $120~{\rm cm}$. The ultrafast response mechanism probably provides opportunities for the dynamic wireless transfer processes, ranging from the travelling electric vehicles to the implantable medical devices.

\emph{Conclusions.}$-$In summary, we proposed two wireless transfer schemes in non-Hermitian QBs with linear and nonlinear $\mathcal{PT}$ symmetries to extract the useful energy stored in QBs. We mainly explored the energy transfer processes from non-Hermitian QBs to consumption-hubs, and the effects of the variable separation distance between the battery and the consumption-hub on the wireless transfer schemes. For the linear gain, the transfer energy performs pseudo-closed oscillations in the unbroken region and grows monotonously in other regions. For the nonlinear gain, the transfer energy eventually reach and remain steady-state values in the total regions under the feedback mechanism of the nonlinear saturable gain. The results reveal that the significant robustness of the static wireless transfer and the ultrafast response of the dynamic wireless transfer with the sudden movements between $20~{\rm cm}$ to $120~{\rm cm}$. In Supplemental Material \cite{supplementalmaterial}, we also discussed counterintuitive advantages of the storage energy and simulated some peculiar properties of the non-Hermitian QBs on Simulation Program with Integrated Circuit Emphasis \cite{ltspice}. Our work overcomes energy bottlenecks for wireless transfer schemes in QBs and may provide opportunities towards real-world applications ranging from wireless charging of electric vehicles to wireless powering of medical devices.
%\nocite{*}

The authors thank Wen-Lei Zhao and Jia-Ming Zhang for their helpful discussions. The work is supported by the National Natural Science Foundation of China (Grants No.~12475026 and No.~12075193) and the Natural Science Foundation of Gansu Province (No.~25JRRA799).
\bibliography{refercence}

\onecolumngrid
\newpage
\section{Supplemental Material for ``Non-Hermitian quantum battery''}
\section{\uppercase\expandafter{\romannumeral 1}. Derivation of the transfer energy}
The task of finding the analytic solution of the transfer energy can reduce to the problem of solving the evolution equations
\begin{equation}
  i\frac{d}{dt}
  \begin{pmatrix}
  \psi_{A}\\\psi_{B}
  \end{pmatrix}
  =H
  \begin{pmatrix}
  \psi_{A}\\\psi_{B}
  \end{pmatrix}.
\end{equation}
For a linear gain $g=\gamma$, the equations become
\begin{eqnarray}
\begin{aligned}
&\dot{\psi_{A}}=(-i\omega_{0}+\gamma)\psi_{A}-i\kappa\psi_{B},\\
&\dot{\psi_{B}}=(-i\omega_{0}-\gamma)\psi_{B}-i\kappa\psi_{A},
\end{aligned}
\end{eqnarray}
the relationship between two field amplitudes follows
\begin{equation}
\psi_{A}=\frac{i}{\kappa}\dot{\psi_{B}}-\frac{\omega_{0}-i\gamma}{\kappa}\psi_{B}.
\end{equation}
The field amplitude in the right resonator satisfies
\begin{equation}
\ddot{\psi_{B}}+i2\omega_{0}\dot{\psi_{B}}+(\kappa^{2}-\omega_{0}^{2}-\gamma^{2})\psi_{B}=0,
\end{equation}
and the roots of its characteristic equation are
\begin{equation}
\lambda_{\pm}=-i\omega_{0}\pm\sqrt{\gamma^{2}-\kappa^{2}}.
\end{equation}

In the unbroken region $|\gamma|<|\kappa|$ and the broken region $|\gamma|>|\kappa|$, the characteristic equation has two unequal roots and the solution of the differential equation is
\begin{equation}
\psi_{B}=C_{+}{\rm e}^{\lambda_{+}t}+C_{-}{\rm e}^{\lambda_{-}t},
\end{equation}
substituting the initial conditions of the field amplitudes $\psi_{A}(0)=1,~\psi_{B}(0)=0$ and we can obtain
\begin{equation}
C_{+}=-C_{-}=-\frac{i\kappa}{2\sqrt{\gamma^{2}-\kappa^{2}}}.
\end{equation}
The field amplitude in the right resonator reads
\begin{equation}
\psi_{B}=
\begin{cases}
\frac{-\kappa\left(\sin\omega_{0}t+i\cos\omega_{0}t\right)}{\sqrt{\kappa^{2}-\gamma^{2}}}\sin\left(\sqrt{\kappa^{2}-\gamma^{2}}t\right),&|\gamma|<|\kappa|,
\\ \\
\frac{-\kappa\left(\sin\omega_{0}t+i\cos\omega_{0}t\right)}{\sqrt{\gamma^{2}-\kappa^{2}}}\sinh\left(\sqrt{\gamma^{2}-\kappa^{2}}t\right),&|\gamma|>|\kappa|.
\end{cases}
\end{equation}

At the EP for $|\gamma|=|\kappa|$, the two roots become equal and the solution is given by
\begin{equation}
\psi_{B}=\left(C_{+}+C_{-}t\right){\rm e}^{\lambda_{+}t},
\end{equation}
the constants under the same initial conditions can be calculated as
\begin{equation}
C_{+}=0,~C_{-}=-i\kappa.
\end{equation}
The field amplitude at the EP is
\begin{equation}
\psi_{B}=-\kappa t(\sin\omega_{0}t+i\cos\omega_{0}t),~|\gamma|=|\kappa|.
\end{equation}

Given all that, the transfer energy defined in Eq.~(\ref{energy}) is as follows
\begin{equation}
\label{analyticenergy}
E=\omega_{0}|\psi_{B}|^{2}=
\begin{cases}
\omega_{0}\kappa^{2}t^{2},&|\gamma|=|\kappa|,\\ \\
\frac{\omega_{0}\kappa^{2}}{\kappa^{2}-\gamma^{2}}\sin^{2}\left(\sqrt{\kappa^{2}-\gamma^{2}}t\right),&|\gamma|<|\kappa|,\\ \\
\frac{\omega_{0}\kappa^{2}}{\gamma^{2}-\kappa^{2}}\sinh^{2}\left(\sqrt{\gamma^{2}-\kappa^{2}}t\right),&|\gamma|>|\kappa|.
\end{cases}
\end{equation}
\section{\uppercase\expandafter{\romannumeral 2}. Derivation of the storage energy}
\begin{figure}[htbp]
\centering
\includegraphics[width=0.47\textwidth]{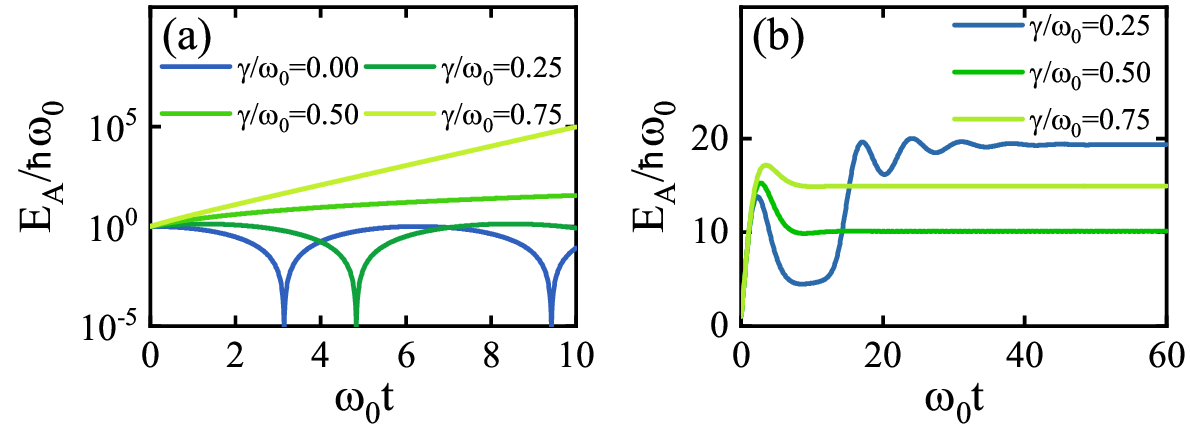}
\caption{The storage energy of non-Hermitian QBs with linear and nonlinear gains. The storage energy has the same advantages as the transfer energy although the battery has small initial energy and the parameters are the same as those in Fig.~\ref{fig2} and Fig.~\ref{fig3}.}
\label{fig5}
\end{figure}

In order to observe the energy stored in non-Hermitian QBs during wireless transfer processes, we introduce the storage energy defined as the energy stored in the left resonator
\begin{equation}
E_{A}=\omega_{0}|\psi_{A}|^{2},
\end{equation}
and we can obtain the analytical expression of the storage energy follows as
%\begin{equation}
%\psi_{A}=
%\begin{cases}
%\left(1+\gamma t\right){\rm e}^{-i\omega_{0}t},&|\gamma|=|\kappa|,
%\\ \\
%\left(\frac{\gamma}{\sqrt{\kappa^{2}-\gamma^{2}}}\sin\left(\sqrt{\kappa^{2}-\gamma^{2}}t\right)+\cos\left(\sqrt{\kappa^{2}-\gamma^{2}}t\right)\right){\rm e}^{-i\omega_{0}t},&|\gamma|<|\kappa|,
%\\ \\
%\left(\frac{\gamma}{\sqrt{\gamma^{2}-\kappa^{2}}}\sinh\left(\sqrt{\gamma^{2}-\kappa^{2}}t\right)+\cosh\left(\sqrt{\gamma^{2}-\kappa^{2}}t\right)\right){\rm e}^{-i\omega_{0}t},&|\gamma|>|\kappa|.
%\end{cases}
%\end{equation}
\begin{equation}
E_{A}=
\begin{cases}
\omega_{0}\left(1+\gamma t\right)^2,&|\gamma|=|\kappa|,
\\
\omega_{0}\left(\frac{\gamma}{\sqrt{\kappa^{2}-\gamma^{2}}}\sin\left(\sqrt{\kappa^{2}-\gamma^{2}}t\right)+\cos\left(\sqrt{\kappa^{2}-\gamma^{2}}t\right)\right)^2,&|\gamma|<|\kappa|,
\\
\omega_{0}\left(\frac{\gamma}{\sqrt{\gamma^{2}-\kappa^{2}}}\sinh\left(\sqrt{\gamma^{2}-\kappa^{2}}t\right)+\cosh\left(\sqrt{\gamma^{2}-\kappa^{2}}t\right)\right)^2,&|\gamma|>|\kappa|.
\end{cases}
\end{equation}
Figure~\ref{fig5} illustrates the dynamic evolutions of the storage energy in non-Hermitian QBs for different loss rates. The storage energy shows pseudo-closed oscillations and monotonous growths in linear cases while reach and remain steady-state values in nonlinear cases. These counterintuitive advantages arise from the fact that non-Hermitian QBs with gains play the role of sources that can persistently extract energy from environment and store it in batteries or transfer it to consumption-hubs, thus perfectly conform law of conservation of energy. Compared to previous wireless schemes, in which the ideal energy stored in QBs performs Rabi-like oscillations induced by memory effects in non-Markovian environments \cite{PhysRevA.102.052223,PhysRevLett.132.090401}, the advantages of the storage energy in our schemes can break Rabi-like energy bottlenecks and exhibit considerable efficiency and significant stability.
\section{\uppercase\expandafter{\romannumeral 3}. Derivation of the coupled-mode equations}
To find out the quantitative relation between the physical quantities and the electronic elements, next we will derive the dynamic equations of the non-Hermitian system from the electronic circuit as depicted in Fig.~\ref{fig1}~(b). The circuit consists of two LC resonators, where the left resonator exhibits nonlinear gain realized by an effective negative resistor while the right resonator exhibits loss arising from a normal resistor. According to the magnetic induction and the Kirchhoff's laws, the voltages on the red nodes satisfy
\begin{eqnarray}
\begin{aligned}
\label{voltages}
&U_{A}=-i\omega\left(L_{A}I_{A}-MI_{B}\right),\\
&U_{B}=-i\omega\left(L_{B}I_{B}-MI_{A}\right),
\end{aligned}
\end{eqnarray}
and the currents flowing through two inductors follow
\begin{eqnarray}
\begin{aligned}
\label{currents}
&I_{A}=i\omega C_{A}U_{A}+\frac{U_{A}}{R_{A}},\\
&I_{B}=i\omega C_{B}U_{B}-\frac{U_{B}}{R_{B}}.
\end{aligned}
\end{eqnarray}
Here we suppose that the entire circuit is working with a time-harmonic field ${\rm e}^{-i\omega t}$. When the currents in Eq.~(\ref{currents}) are replaced by the currents ascertained in Eq.~(\ref{voltages}), the equations are reformulated as
\begin{eqnarray*}
\begin{aligned}
&\left[\frac{\omega_{A}^{2}}{2}-\frac{k\omega^{2}}{2}+\frac{ik\omega}{2C_{A}R_{A}}\right]U_{A}+\frac{\omega_{A}^{2}}{2}\frac{M}{L_{B}}U_{B}=0,\\
&\left[\frac{\omega_{B}^{2}}{2}-\frac{k\omega^{2}}{2}-\frac{ik\omega}{2C_{B}R_{B}}\right]U_{B}+\frac{\omega_{B}^{2}}{2}\frac{M}{L_{A}}U_{A}=0,
\end{aligned}
\end{eqnarray*}
where $k=1-M^{2}/(L_{A}L_{B})$, $\omega_{A,B}=1/\sqrt{L_{A,B}C_{A,B}}$, and $M$ is the mutual inductance. Assume that the frequency $\omega$ of the time-harmonic field is nearly resonant with the frequency $\omega_{A,B}$ of each resonators and ignore the second-order term $M^{2}/(L_{A}L_{B})$ since the mutual inductance is small compared with two inductances. The equations are reduced to
\begin{eqnarray}
\begin{aligned}
&\left(\omega_{A}-\omega+\frac{i}{2C_{A}R_{A}}\right)U_{A}+\frac{\omega_{A}}{2}\frac{M}{L_{B}}U_{B}=0,\\
&\left(\omega_{B}-\omega-\frac{i}{2C_{B}R_{B}}\right)U_{B}+\frac{\omega_{B}}{2}\frac{M}{L_{A}}U_{A}=0.
\end{aligned}
\end{eqnarray}
The above equations can be rewritten in matrix form as
\begin{equation}
  \begin{pmatrix}
  \omega_{A}-\omega+\frac{i}{2C_{A}R_{A}}&\frac{\omega_{B}}{2}\frac{M}{L_{A}}\\
  \\
  \frac{\omega_{A}}{2}\frac{M}{L_{B}}&\omega_{B}-\omega-\frac{i}{2C_{B}R_{B}}
  \end{pmatrix}
  \begin{pmatrix}
  U_{A}\\U_{B}
  \end{pmatrix}
  =0.
\end{equation}
For simplicity, we map the Kirchhoff's equations into the coupled-mode equations
\begin{equation}
  \begin{pmatrix}
  \omega_{A}+ig&\kappa\\
  \kappa&\omega_{B}-i\gamma
  \end{pmatrix}
  \begin{pmatrix}
  U_{A}\\U_{B}
  \end{pmatrix}
  =\omega
  \begin{pmatrix}
  U_{A}\\U_{B}
  \end{pmatrix},
\end{equation}
where the resonant frequency $\omega_{0}=\omega_{A,B}$, the self inductance $L=L_{A,B}$, the capacitance $C=C_{A,B}$, the coupling strength $\kappa=\omega_{0}M/(2L)$, the gain rate $g=1/(2CR_{A})$, and the loss rate $\gamma=1/(2CR_{B})$.

\begin{figure*}[htbp]
\centering
\includegraphics[width=0.95\textwidth]{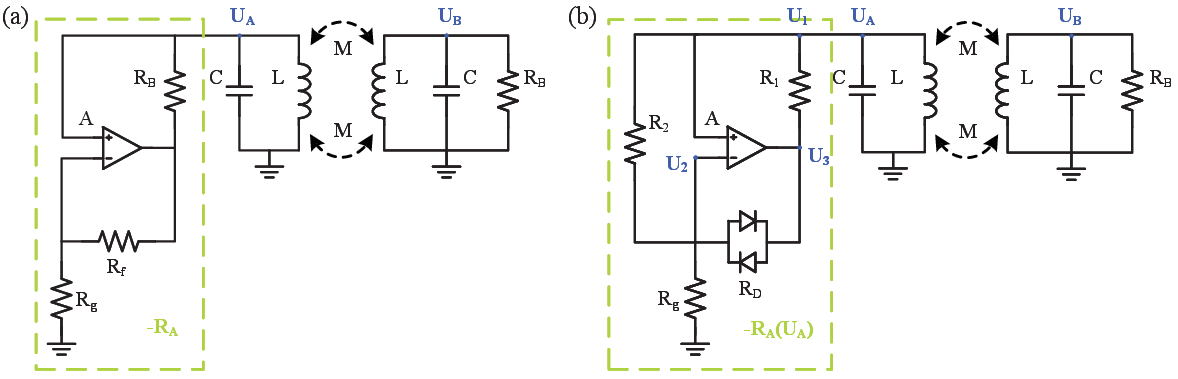}
\caption{The effective circuit diagrams for linear (left) and nonlinear (right) $\mathcal{PT}$-symmetric transfer schemes, including capacitors (C), inductors (L), resistors (R), amplifiers (A) and diodes (D). The green dashed boxes mark the effective negative resistors provided by voltage-doubling buffers.}
\label{fig6}
\end{figure*}
\section{\uppercase\expandafter{\romannumeral 4}. Derivation of the linear and nonlinear gains}
The dashed box shown in Fig.~\ref{fig6}~(b) illustrates the effective circuit diagram of the negative resistor, which is composed by an operational amplifier, two diodes and three normal resistors. In an ideal amplifier, the voltage on non-inverting input is equal to the voltage on inverting input and the input impedances are so large that almost no currents flow into the inputs. The voltage on each node is given by
\begin{equation}
U_{1}=U_{2}=U_{A},~U_{3}=\frac{R_{D}}{R_{g}}U_{A}+U_{A}.
\end{equation}
where $R_{g}$ and $R_{D}$ represent the ground resistor and the composite resistor of the parallel diodes. The currents flowing out of the node 1 are
\begin{equation}
I_{R_{1}}=\frac{U_{1}-U_{3}}{R_{1}}=-\frac{R_{D}}{R_{1}R_{g}}U_{A},~I_{R_{2}}=\frac{1}{R_{2}}U_{A}.
\end{equation}
The current passing through the node 1 follows the Kirchhoff's laws and can be expressed as
\begin{equation}
-\frac{U_{A}}{R_{A}}=-\frac{R_{D}}{R_{1}R_{g}}U_{A}+\frac{U_{A}}{R_{2}},
\end{equation}
where the effective negative resistor inside the dashed box is equivalent to
\begin{equation}
-R_{A}=\frac{R_{1}R_{2}R_{g}}{-R_{2}R_{D}+R_{1}R_{g}}.
\end{equation}
The nonlinear gain in the left LC resonator is
\begin{equation}
g=\frac{1}{2CR_{A}}=\frac{R_{D}}{2CR_{1}R_{g}}-\frac{1}{2CR_{2}}.
\end{equation}
It is worth noting that the properties of the diodes result in the monotonic decreasing of the resistor $R_{D}$ as a function of the voltage $U_{A}$. Thus, we conclude that the nonlinearity of the gain arises from the parallel diodes in the electronic circuit. Similarly, the effective negative resistor shown in Fig.~\ref{fig6}~(a) follows
\begin{equation}
-R_{A}=-R_{B}R_{g}/R_{f},
\end{equation}
and under the condition $R_{f}=R_{g}$, the linear gain can be written as
\begin{equation}
g=\frac{1}{2CR_{A}}=\frac{1}{2CR_{B}}=\gamma.
\end{equation}

\begin{figure}[htbp]
\centering
\includegraphics[width=0.47\textwidth]{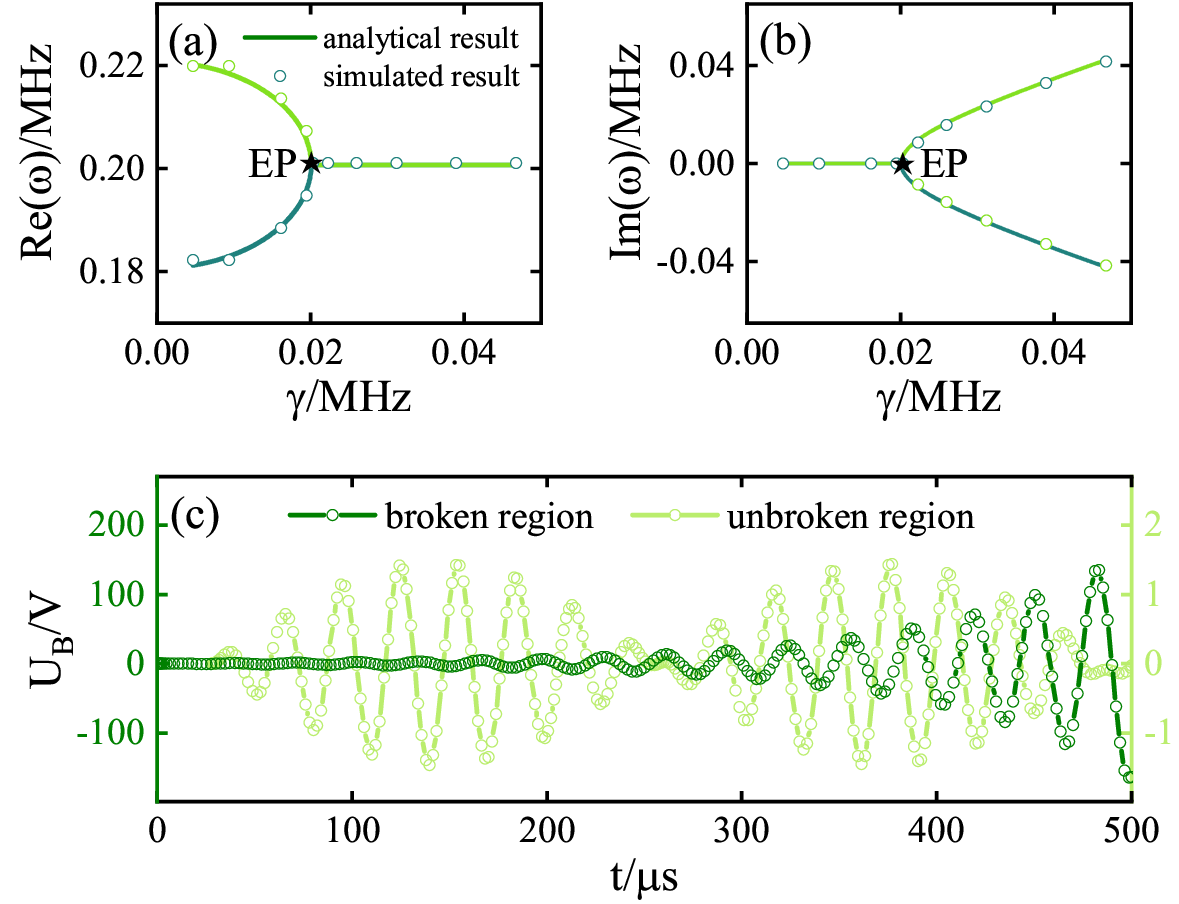}
\caption{The simulation of the non-Hermitian QB with linear gain on LTspice. (a),~(b) The comparison of simulation and analytical results of the eigenfrequencies $\omega$. (c) The time evolutions of the voltages $U_{B}$ in unbroken and broken regions. The initial conditions used in the linear $\mathcal{PT}$-symmetric circuit are $U_{A}(0)=1{\rm{V}}$ and $U_{B}(0)=0{\rm{V}}$. Other parameters are set as $M/L=0.2, L=2.32{\rm{mH}}, C=10.7{\rm{nF}}, R_{B}=2{\rm{k\Omega}}, 3{\rm{k\Omega}}$, and $R_{f}=R_{g}=1{\rm{k\Omega}}$.}
\label{fig7}
\end{figure}
\section{\uppercase\expandafter{\romannumeral 5}. Observation of the EP and the nonlinear gain on LTspice}
As depicted in Fig.~\ref{fig6}~(a), a linear $\mathcal{PT}$-symmetric circuit consists of two magnetically coupled LC resonators with loss and gain elements, which arises from the normal resistor and the effective negative resistor provided by a voltage-doubling buffer, respectively. To observe the EP mentioned above, we also perform time-domain circuit simulations on Simulation Program with Integrated Circuit Emphasis (SPICE) \cite{ltspice}. The results reveal that the oscillation frequencies of voltage waveforms and the amplification (attenuation) degrees of voltage amplitudes are associated with the real and imaginary parts of the eigenfrequencies, where the EP is marked by black pentagrams in Fig.~\ref{fig7}~(a) and (b). In addition, the most interesting behaviors of voltages appear around the EP, that is, the voltage envelope oscillates periodically in the unbroken region and grows exponentially in the broken region.

\begin{figure}[htbp]
\centering
\includegraphics[width=0.47\textwidth]{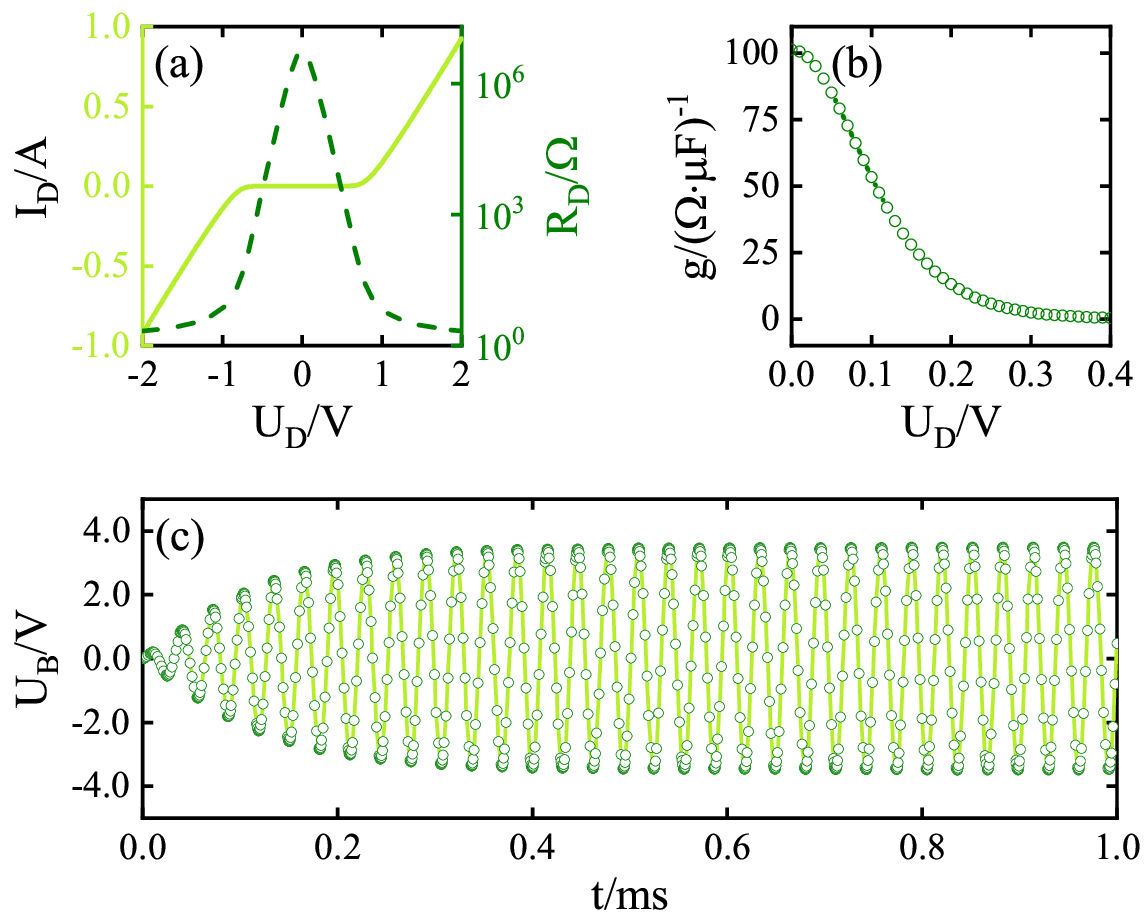}
\caption{The simulation of the non-Hermitian QB with nonlinear gain on LTspice. (a) The volt-ampere characteristic curve and the effective resistance $R_{D}$ of two parallel diodes. (b) The nonlinear gain $g$ arises from the parallel diodes in the nonlinear $\mathcal{PT}$-symmetric circuit. (c) The time evolution of the voltage $U_{B}$ in the left LC resonator. The resistors in the nonlinear circuit are set as $R_{1}=500{\rm{\Omega}}, R_{2}=10{\rm{k\Omega}}, R_{g}=5{\rm{k\Omega}}$, and $R_{B}=1{\rm{k\Omega}}$. Other parameters and the initial conditions are the same as those in the linear circuit.}
\label{fig8}
\end{figure}
Figure~\ref{fig6}~(b) shows a nonlinear $\mathcal{PT}$-symmetric circuit consisting of two coupled LC resonators with loss and nonlinear gain, where the dashed box marks the nonlinear gain element and the nonlinearity arises from the monotonic decreasing resistance of two parallel diodes. The effective resistance and the nonlinear gain as functions of the voltage on gain side are shown in Fig.~\ref{fig8}~(a),~(b). We perform the simulations with an initial voltage on right resonator, the voltage envelope reaches a steady state at around $0.4~{\rm{ms}}$ and the voltage waveform oscillates as the steady state with eigenfrequencies. In fact, such simulations of non-Hermitian QBs are limited to classical systems in which only the amplitude information, not the phase information, is measured \cite{Naghiloo2019}. Thus, our simulations on LTspice only demonstrate some advantages of non-Hermitian QBs, rather than  implying that QBs can be implemented in electronic circuits.
% Produces the bibliography via BibTeX.

\end{document}